\documentclass{article}
\usepackage[utf8]{inputenc}
\usepackage[margin=1.5in]{geometry}
\usepackage{authblk}
\usepackage{siunitx}
\usepackage{amsfonts}
\usepackage{hyperref}
\usepackage{caption}
\usepackage{subcaption}
\usepackage{graphicx}
\usepackage{makecell}
\graphicspath{ {./figs/} }
\usepackage[style=nature]{biblatex}
\title{Microstructure Generation via Generative Adversarial Network for Heterogeneous, Topologically Complex 3D Materials}

\author[a,b,f]{Tim Hsu}
\author[a,c]{William K. Epting}
\author[a,b]{Hokon Kim}
\author[d,e]{Harry W. Abernathy}
\author[d]{Gregory A. Hackett}
\author[a,b]{Anthony D. Rollett}
\author[a,b]{Paul A. Salvador}
\author[a,b]{Elizabeth A. Holm\thanks{Corresponding Author, eaholm@andrew.cmu.edu}}

\affil[a]{U.S. DOE National Energy Technology Laboratory, Pittsburgh, PA 15236, USA}
\affil[b]{Materials Science and Engineering, Carnegie Mellon University, Pittsburgh, PA 15213, USA}
\affil[c]{Leidos Research Support Team, Pittsburgh, PA 15236, USA}
\affil[d]{U.S. DOE National Energy Technology Laboratory, Morgantown, WV 26505, USA}
\affil[e]{Leidos Research Support Team, Morgantown, WV 26505, USA}
\affil[f]{Lawrence Livermore National Laboratory, Livermore, CA 94550, USA}
\date{}

\begin{document}

\maketitle

\begin{abstract}
Using a large-scale, experimentally captured 3D microstructure dataset, we implement the generative adversarial network (GAN) framework to learn and generate 3D microstructures of solid oxide fuel cell electrodes. The generated microstructures are visually, statistically, and topologically realistic, with distributions of microstructural parameters, including volume fraction, particle size, surface area, tortuosity, and triple phase boundary density, being highly similar to those of the original microstructure. These results are compared and contrasted with those from an established, grain-based generation algorithm (DREAM.3D). Importantly, simulations of electrochemical performance, using a locally resolved finite element model, demonstrate that the GAN generated microstructures closely match the performance distribution of the original, while DREAM.3D leads to significant differences. The ability of the generative machine learning model to recreate microstructures with high fidelity suggests that the essence of complex microstructures may be captured and represented in a compact and manipulatable form.
\end{abstract}

\section{Introduction}
As a component of integrated computational materials engineering (ICME), methods to generate realistic simulation volumes are essential for modeling and simulating materials with complex microstructures \cite{dimiduk2018perspective}. While obtaining microstructures experimentally guarantees realistic simulation volumes, the cost or difficulty of microstructural characterization often limits the size or number of microstructures that can be sampled, particularly in 3D. Thus, to support computational design and performance surveys, a common goal is to synthesize statistically representative sets of microstructural realizations \cite{decost2019vision}. A number of successful approaches have been developed, such as those based on n-point correlation functions \cite{jiao2008modeling, torquato2002random}, ellipsoid packing \cite{mandal2018generation, groeber2014dream}, physical descriptors \cite{xu2014descriptor}, Gaussian random field \cite{jiang2013efficient}, and Markov random field \cite{bostanabad2016stochastic}. Bostanaband et al. \cite{bostanabad2018computational} provide an extensive review of existing microstructure reconstruction techniques. However, despite their successes, most of them require assumptions about the underlying structure, and thus are often limited to either binary (two-phase) microstructures, uniform microstructures consisting of primitive/regular geometries, or a combination of both; there is an intrinsic limit of generality that prohibits the use of these statistical methods for a wide range of heterogeneous, complex material systems.

In recent years, deep learning based generative models, such as variational autoencoders \cite{kingma2013autoencoding} or generative adversarial networks (GANs) \cite{goodfellow2014generative}, have attracted the interests of materials scientists for their potential in microstructure image generation \cite{yang2018microstructural, cang2018improving, iyer2019conditional, singh2018physics}. Relative to other reconstruction methods, generative models based on deep convolutional neural networks make no assumption regarding the underlying structure of the image data. Such models can therefore be vastly general. Additionally, these models have the ability to learn the underlying data distribution of their input dataset. In other words, a well-trained generative model can (theoretically) synthesize an infinite number of unique microstructure realizations from the learned data distribution at relatively trivial speeds and compute costs. 

Most generative models for microstructure focus on 2D representations \cite{yang2018microstructural, cang2018improving, iyer2019conditional, singh2018physics, fokina2019microstructure}. For example, Yang et al. \cite{yang2018microstructural}, Cang et al. \cite{cang2018improving}, and Singh et al. \cite{singh2018physics} successfully applied them to represent 2D binary (two-phase) microstructure images. Iyer et al. \cite{iyer2019conditional} explored the use of a conditional GAN \cite{mirza2014conditional, odena2016conditional} and Fokina et al. \cite{fokina2019microstructure} used a style-based GAN \cite{karras2018stylebased} for capturing complex, multiphase, grayscale 2D microstructure images. There is some prior work focused on 3D microstructure representation: Mosser et al. \cite{mosser2018stochastic, mosser2017reconstruction} utilized a GAN to generate 3D porous media microstructures, including from binary image information \cite{mosser2017reconstruction} and from uniform microstructures made of closely packed spherical grains (i.e., simple or regular geometries) \cite{mosser2018stochastic}. Most recently, Gayon-Lombardo et al. \cite{gayonlombardo2020pores} applied GANs to generate 3D, three-phase microstructural volumes of moderate size (or $64 \times 64 \times 64$ voxels) with periodic boundary conditions and good agreement with the target structures. These initial works confirm the general utility of generative models in representing microstructures.

\begin{figure}
    \centering
    \includegraphics[width=\textwidth]{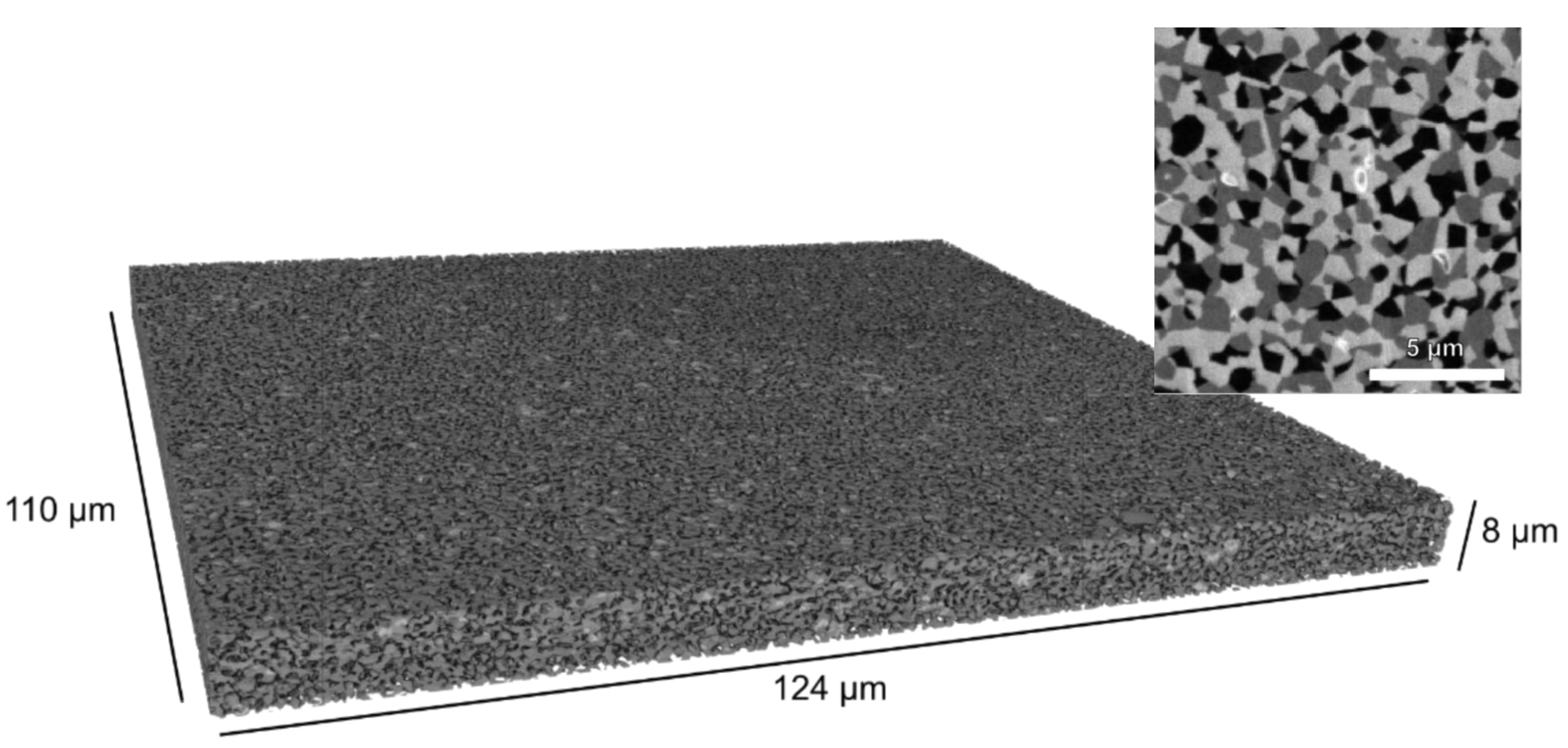}
    \caption{Complex, multiphase microstructure of a commercial solid oxide fuel cell (SOFC) anode containing yttria-stabilized zirconia (YSZ, light gray), nickel (Ni, dark gray), and pore phases (transparent). The $110 \times 124 \times 8$ \si{\um^3} 3D volume was obtained by PFIB-SEM; bright spots are imaging artifacts. The inset shows an example 2D cross-section, imaged by SEM, where YSZ is bright gray, Ni is dark gray, and pores are black.
}
    \label{fig:SOFC-original}
\end{figure}

In this paper, we demonstrate that the GAN framework with deep convolutional layers \cite{radford2015unsupervised} can learn and generate grayscale, heterogeneous, complex multiphase microstructures in 3D at a scale appropriate for simulation of material properties. Our model material is a three-phase commercial solid oxide fuel cell (SOFC) anode, as shown in Figure~\ref{fig:SOFC-original}. The original microstructure image data was experimentally captured using Xe plasma focused ion beam combined with scanning electron microscope (Xe PFIB-SEM) \cite{mahbub2017method, hsu2018mesoscale} and represents the largest 3D volume of SOFC material characterized to date. The synthetic microstructures produced by the GAN model are both visually and statistically realistic when compared with the experimental microstructure and with microstructures synthesized by an established, grain-based generation algorithm (DREAM.3D) \cite{groeber2014dream, mandal2018generation}. Importantly, simulations of electrochemical performance, using a locally resolved finite element model \cite{hsu2020high, hsu2020Moose}, demonstrate that the GAN generated microstructures closely match the performance distribution of the real SOFC material. The ability of the machine learning model to recreate microstructures with high fidelity makes it a valuable addition to the ICME tool set.

\section{Methods}
\subsection{Experimental Microstructure Data Set}
As shown in Figure~\ref{fig:SOFC-original}, the SOFC anode microstructure is a three-phase, 3D composite of highly interconnected yttria-stabilized zirconia (YSZ), nickel (Ni), and pores. This material comprises sintered grains with characteristic length scales on the order of 0.5 \si{\um} and exhibits substantial variability in volume fractions over the length scale of 15 \si{\um} (or 30 times the characteristic length scale) \cite{hsu2018mesoscale, mahbub2017method}. The extent and connectivity of triple phase boundaries (TPB), where YSZ, Ni, and pore phases meet in a line junction, governs the performance of the material. The TPB network topology is complex and interconnected, necessitating a 3D microstructural representation for accurate performance modeling.

The microstructure training dataset originates from the active anode layer of a commercial SOFC sample supplied by Materials and Systems Research, Inc. (Salt Lake City, UT), which was imaged using PFIB-SEM as previously described \cite{mahbub2017method, hsu2018mesoscale}. The resulting microstructure data is a large-scale, 3D volumetric grayscale image whose dimension is $1697 \times 1900 \times 124$ voxels, or $110 \times 124 \times 8$ \si{\um^3}, thus the voxel size is $65 \times 65 \times 65$ \si{\nm^3}. Each phase images with a characteristic grayscale intensity, and microscopy artifacts may appear as well. Subvolumes of this microstructure dataset are sampled during GAN training sessions. 

\subsection{DREAM.3D Synthetic Microstructures}
As described elsewhere \cite{hsu2018mesoscale, mahbub2020quantitative}, synthetic microstructures were constructed using the DREAM.3D code package \cite{groeber2014dream} (BlueQuartz Software, Springboro, OH), which generates multiphase microstructures via an ellipsoid packing scheme. The parameters specified were phase fraction and the first and second moments of the log-normal particle size distribution, which were chosen to match as closely as possible the measured vales for each phase in the anode microstructure. Forty distinct microstructures of $12.48 \times 12.48 \times 12.48$ \si{\um^3} were first generated, and they were then split into octants to produce 320 subvolumes of $6.24 \times 6.24 \times 6.24$ \si{\um^3} ($96 \times 96 \times 96$ voxels).

\subsection{GAN Implementation}
Convolutional neural networks (CNNs) are a class of artificial neural networks particularly well-suited to image data and computer vision applications \cite{szeliski2010cv}. In its forward mode, a CNN takes in an image and applies a series of signal processing steps (typically filter convolutions, rectification operations, and pooling or downsampling) in order to generate a compact representation of the visual information contained in the image, termed the feature vector. The feature vector can then be used for a variety of image processing tasks, including visual similarity and classification \cite{decost2017manifold}. The architecture of the CNN, including the type, number, order, and connectivity of the operations, is user-defined, and the training process attempts to optimize the many internal variables for best performance.

\begin{figure}
    \centering
    \includegraphics[width=\textwidth]{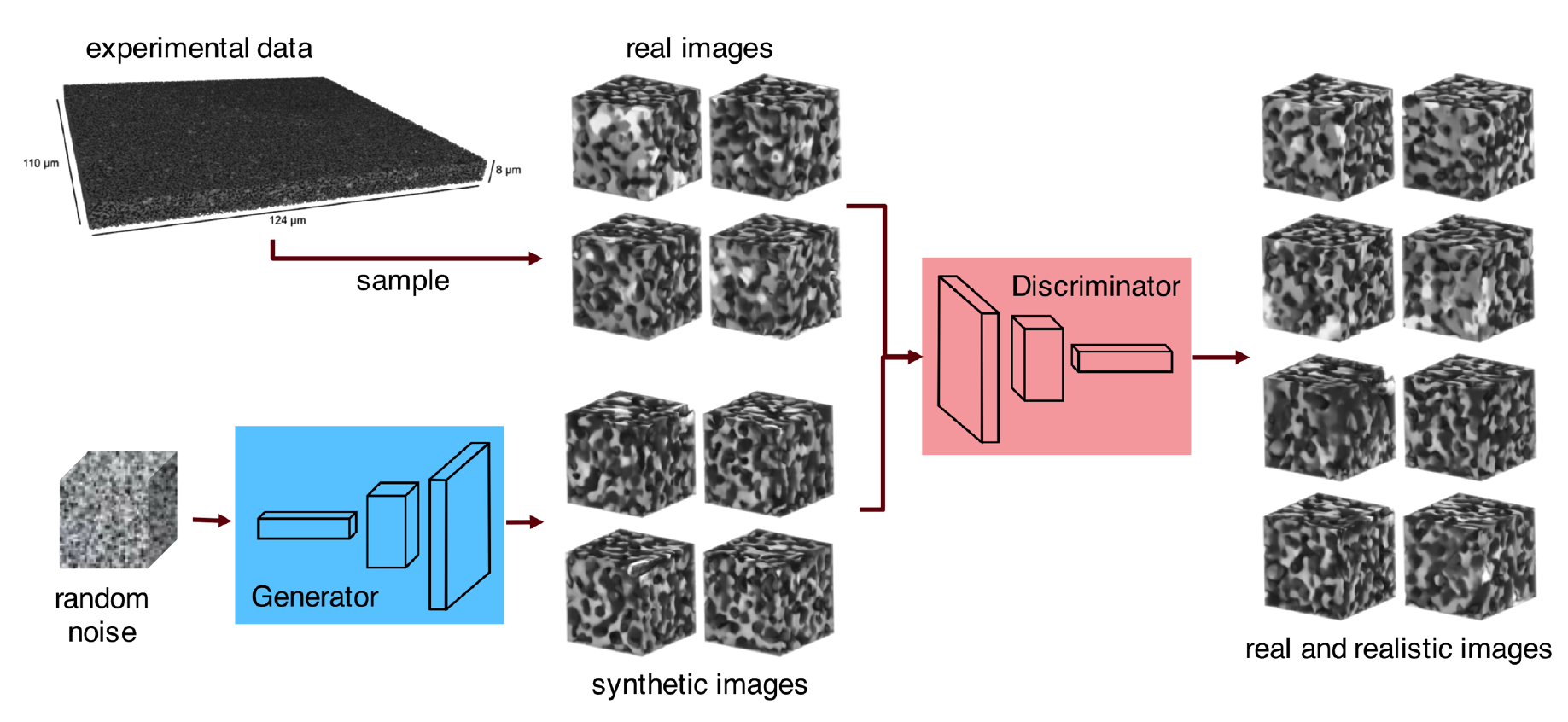}
    \caption{Schematic of the GAN of the model. The generator transforms a random noise vector $z$ into a 3D image, and the discriminator classifies a 3D image as real or generated. The discriminator is trained using a combination of real microstructural volumes, sampled from the experimental data, and fake images generated by the generator network. The generator is trained based on the success/failure of previously generated images. In a fully trained GAN, the discriminator will approve both real images and sufficiently realistic generated images.}
    \label{fig:GAN-diagram}
\end{figure}

CNNs can also operate in reverse, taking in a feature vector and transforming it into an image; this is the same basic concept as an auto-encoder \cite{kingma2014adam}. As shown in Figure~\ref{fig:GAN-diagram}, a generative adversarial network (GAN) pits two CNNs against each other: The generator, in reverse mode, creates candidate images, which the discriminator (or critic), in forward mode, attempts to classify as generated/fake. The discriminator is trained to recognize real images (in our case, original microstructures). Both networks participate in a feedback loop, so the generator learns to make more convincing fakes, and the discriminator learns to spot them. When the generator creates images of sufficient quality to consistently fool the discriminator, training is complete and the GAN may be used to generate realistic synthetic images \cite{goodfellow2014generative}.

We implemented the GAN model using the Tensorflow platform \cite{abadi2016tensorflow} with the Keras API (\url{https://keras.io/}). The model architecture, given in Table~\ref{tab:GAN-table}, largely follows the work by Miyato et al. \cite{miyato2018spectral} on spectral normalization and is specifically based on their standard architecture. However, for the generator, we used upsampling and convolutional layers, rather than standard deconvolutional (transposed convolution) layers, to avoid potential checkerboard artifacts \cite{odena2016deconvolution}. We used a public code repository at \url{https://github.com/IShengFang/SpectralNormalizationKeras} to implement spectral normalization applied to convolutional and dense layers in the critic network. The Adam optimizer \cite{kingma2014adam} was used for training the model. Relevant hyperparameters are shown in Table~\ref{tab:hyperparams}.

During training, for each discriminator update, a batch of subvolumes randomly sampled from the experimental 3D microstructure and a batch of 3D images randomly generated from the generator are fed into the discriminator network to compute the Wasserstein loss \cite{arjovsky2017wasserstein} and the subsequent gradient update. The sampling process involves cropping and symmetry operations (rotation and mirror-flipping) of the cropped subvolumes to increase the perceived training data variability for the critic. 

We trained the model in a distributed fashion with 64 high-performing GPUs (Nvidia Tesla P100 PCIe) on a supercomputer (NETL Joule) using the Horovod framework \cite{sergeev2018horovod}. The training took about 39 hours, or 32,565 iterations of generator update. Due to the nature of the synchronous distributed training model implemented by Horovod, the effective minibatch size is the original minibatch size multiplied by the number of GPU workers, or $8 \times 64 = 512$. Although Goyal et al. \cite{goyal2017accurate} have shown that under synchronous distributed training, the learning rate can be scaled linearly with respect to the minibatch size without accuracy loss for classification tasks using ResNet-50 \cite{he2015deep}, we have observed that the GAN training dynamic is highly sensitive to adjustments made to the learning rate. Therefore, we did not scale the learning rate in order to preserve training stability.

\begin{table}
  \centering
    \caption{Architecture of the GAN model for 3D microstructure generation. All convolutional layers use the ``same'' padding. The slope for all leaky ReLU (lReLU) activations is 0.1. Upsample and Reshape operations are not associated with kernel size, stride, batch normalization, and activation function. Batch normalization operation is applied after activation function, not before.}
    \begin{tabular}{c|c|c|c|c|c|c}
    \hline
        \thead{Layer} &
        \thead{Operation(s)} &
        \thead{Output \\ dimensions} &
        \thead{Kernel \\ size} &
        \thead{Stride} &
        \thead{Batch \\ normalization} &
        \thead{Activation \\ function}\\
    \hline
        \multicolumn{7}{l}{Generator: Input = $z \in \mathbb{R}^{100} \sim \mathcal{N}(0,1)$}\\
    \hline
        $\mathrm{G}_\mathrm{in}$ & 
             \makecell{Dense              \\ Reshape}&  (6,  6,  6, 512) & - & - & -   & Linear \\ \hline
        G1 & \makecell{$2\times$ Upsample \\ Conv3D} & (12, 12, 12, 256) & 4 & 1 & Yes & ReLU\\ \hline
        G2 & \makecell{$2\times$ Upsample \\ Conv3D} & (24, 24, 24, 128) & 4 & 1 & Yes & ReLU\\ \hline
        G3 & \makecell{$2\times$ Upsample \\ Conv3D} & (48, 48, 48,  64) & 4 & 1 & Yes & ReLU\\ \hline
        G4 & \makecell{$2\times$ Upsample \\ Conv3D} & (96, 96, 96,  32) & 4 & 1 & Yes & ReLU\\ \hline
        G5 & \makecell{                      Conv3D} & (96, 96, 96,   1) & 3 & 1 &  No & Tanh\\
  \hline
    \multicolumn{7}{l}{Discriminator: Input = grayscale image $x \in \mathbb{R}^{96\times96\times96\times1}$ }\\
  \hline
       D1 & \makecell{ConvSN3D \\ ConvSN3D} & (48, 48, 48,  32) & \makecell{3 \\ 4} & \makecell{1 \\ 2} & \makecell{No \\ No} & \makecell{lReLU \\ lReLU}\\ \hline
       D2 & \makecell{ConvSN3D \\ ConvSN3D} & (24, 24, 24,  64) & \makecell{3 \\ 4} & \makecell{1 \\ 2} & \makecell{No \\ No} & \makecell{lReLU \\ lReLU}\\ \hline
       D3 & \makecell{ConvSN3D \\ ConvSN3D} & (12, 12, 12, 128) & \makecell{3 \\ 4} & \makecell{1 \\ 2} & \makecell{No \\ No} & \makecell{lReLU \\ lReLU}\\ \hline
       D4 & \makecell{ConvSN3D \\ ConvSN3D} &  (6,  6,  6, 256) & \makecell{3 \\ 4} & \makecell{1 \\ 2} & \makecell{No \\ No} & \makecell{lReLU \\ lReLU}\\ \hline
       D5 & \makecell{ConvSN3D}             &  (6,  6,  6, 512) & \makecell{3}      & \makecell{1}      & No                  & lReLU                    \\ \hline
       $\mathrm{D}_\mathrm{out}$ & DenseSN  & (1) & - & - & - & Linear \\
\hline
   \end{tabular}
 \label{tab:GAN-table}
\end{table}

\begin{table}
    \caption{Model hyperparameters.
             Here, one epoch is equivalent to the critic trained over 100,000 images}
    \centering
    \begin{tabular}{l l l}
        \hline
        Symbol        & Description                                   & Value  \\
        \hline
        $\alpha$      & Learning rate for both generator and critic (initial)  & 0.00005 \\
        $\alpha$      & Learning rate for both generator and critic (after 27 epochs)  & 0.00001 \\
        $\alpha$      & Learning rate for both generator and critic (after 78 epochs)  & 0.000005 \\
        $\alpha$      & Learning rate for both generator and critic (after 147 epochs)  & 0.000001 \\
        $\beta_1$     & First decay rate for Adam optimizer           & 0.1 \\
        $\beta_2$     & Second decay rate for Adam optimizer          & 0.9 \\
        $n_{dis}$     & Number of critic update(s) per generator update  & 1 \\
        $\gamma_{BN}$ & Batch normalization momentum term             & 0.8 \\
        \hline
    \end{tabular}
    \label{tab:hyperparams}
\end{table}

\subsection{Segmentation}\label{sec:segmentation}
Experimental and GAN synthetic grayscale microstructures were segmented into three phases using a custom watershed-based segmentation code, which could be implemented in a high-throughput fashion; the major steps are as follows:

\begin{enumerate}
    \item Sobel filters were applied to an input grayscale image to approximate image gradients.
    \item A 2D density map was generated, plotting the grayscale intensity in voxels (x-axis) versus the gradient between voxels (y-axis).
    \item Three high-density regions in the density map were then manually bounded to indicate what voxels would be selected for seeds in the watershed algorithm. For example, voxels with gradients lower than 1 and intensity values lower than 40 would be selected as seeds for phase 1. This manual step was done once, incorporating the entire experimental data set in the selection process. The same bounds were then used for all 646 sub-volumes during batch segmentation.
    \item A watershed dilation algorithm was then applied using the 3 groups of voxel markers determined above. The algorithm accounts for image gradients while dilating/growing the markers (seeds). The output of this step is a segmented image with three phases. When using the same groups of markers in the commercial AVIZO package, statistically identical results were produced, which we take as validation that the code operates as intended.
\end{enumerate}

As described in our earlier works \cite{epting2012resolving, litster2013morphological, epting2017quantifying}, custom in-house Python codes were used to calculate relevant microstructural properties for each subvolume, including: volume fractions; the average and standard deviation of diameter from the volume-weighted size distribution for each phase (based on an inscribed sphere method); interfacial surface area between each pair of phases; geometric tortuosity factors for each phase; and total and active (connected) TPB densities. For each phase in each subvolume, we also calculated the formation factor $K_i$, which represents the ratio between the effective and bulk diffusivity or conductivity in phase $i$. Here we use the fairly common estimation for the formation factor as the ratio between the phase fraction and geometric tortuosity factor: $K_i = \theta_i/\tau_i$ \cite{litster2013morphological}.

\subsection{Electrochemistry Simulations}
Simulations of the electrochemical performance of 90 3D microstructures were performed, as described previously \cite{hsu2020Moose, hsu2020high}. Briefly, the application ERMINE--- \textbf{E}lectrochemical \textbf{R}eactions in \textbf{Mi}crostructural \textbf{Ne}tworks---was used, which is instantiated within MOOSE, the open-source finite element framework developed by Idaho National Laboratory. 30 subvolumes for each of the original, GAN, and DREAM.3D microstructures were subjected to the workflow described elsewhere \cite{hsu2020Moose}, including appending a 4.16 \si{um} YSZ layer as an electrolyte. Morphology-preserving microstructural meshes were generated  using a commercial meshing software (Simpleware ScanIP+FE 7.0, Synopsys, Inc., Mountain View, CA) and used as the simulation domains. All simulations were carried out on the Joule supercomputer (National Energy Technology Laboratory, Morgantown, WV). 120 cores were used in parallel for simulation of any subvolume, and simulations of many individual subvolumes also could be run in parallel on JOULE. The typical time to complete a simulation for a single subvolume, which involved simulations from 0 to 0.4 V overpotential, was approximately 12 minutes. The total clock time for simulation of all 90 subvolumes was less than an hour using Joule.

\section{Results and Discussion}
\subsection{Visual Similarity}
\begin{figure}
    \centering
    \addtolength{\leftskip}{-1.5cm}
    \includegraphics[width=1.2\textwidth]{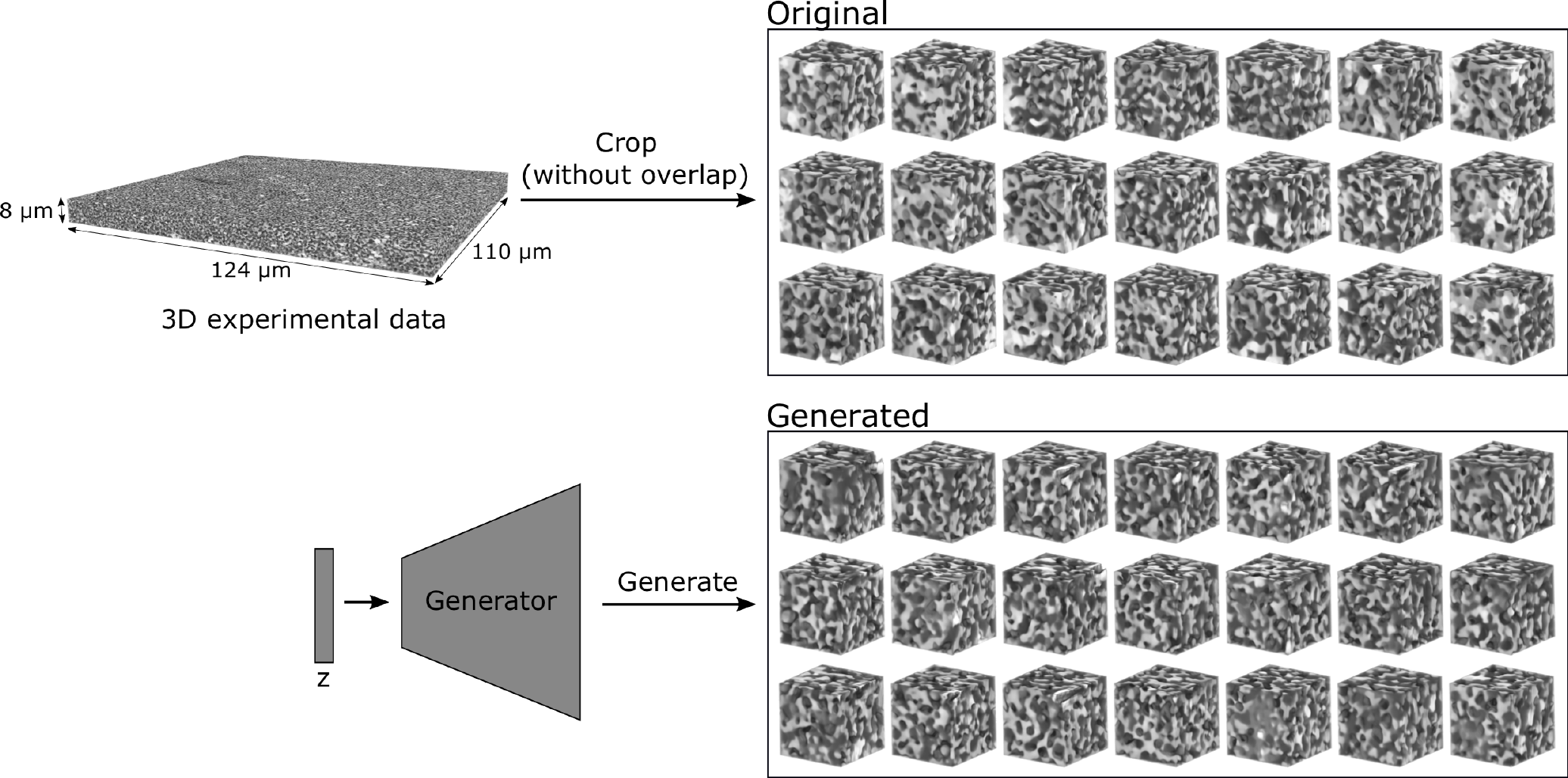}
    \caption{Volume-renderings of representative microstructures cropped from the original experimental data (top) or generated by the trained generator using random noise vector inputs (bottom). Visual similarity between the two sets of images is notable. The cropped/generated volumes have the dimensions of $96 \times 96 \times 96$ voxels, with voxel sizes of $65 \times 65 \times 65$ \si{\nm^3}. Therefore, the cropped volumes have physical dimensions of $6.24 \times 6.24 \times 6.24$ \si{\um^3}.}
    \label{fig:volume-render}
\end{figure}

The 3D geometry and topology of the three-phase SOFC microstructure is highly complex and interconnected, which is by design to have good electrochemical performance \cite{minh1995science, wilson2006three, rolison2009multifunctional, zhang2011three}. This complexity is visually apparent in Figure~\ref{fig:volume-render} (top). A rendering of the whole $124 \times 110 \times 8$ \si{\um^3} volume is given in Figure~\ref{fig:volume-render} (top left), as are 21 grayscale images of $6.24 \times 6.24 \times 6.24$ \si{\um^3} volumes cropped from the original (top right). The grayscale intensities in the experimental data reflect phase contrast, with some microscopy artifacts (e.g., charging). 323 volumes can be cropped without overlap from the experimental data, though the training sets used here include overlaps and data augmentation.

Upon sufficient training, the generator component of the GAN model produces microstructures that visually capture the complexity of the original with striking fidelity. 21 rendered grayscale images produced by the generator are given in Figure~\ref{fig:volume-render} (bottom right). The generator can, in theory, create an infinite number of unique microstructure volumes, and indeed all 21 generated volumes appear to be different. While the GAN-generated structures and original data look very similar, close inspection reveals that the original volumes display more variability than do the generated volumes. For example, some original volumes show strong SEM surface-charging artifacts (which appear as bright white regions), yet the generated volumes contain few or none of these artifacts. This is consistent with observations that GANs tend to reproduce less variation than the data used to train them \cite{karras2017progressive, salimans2016improved}; they learn the average structure more readily than the outliers. Nevertheless, the GAN microstructures appear similar to those in the original distribution.

\begin{figure}
    \centering
    \includegraphics[width=\textwidth]{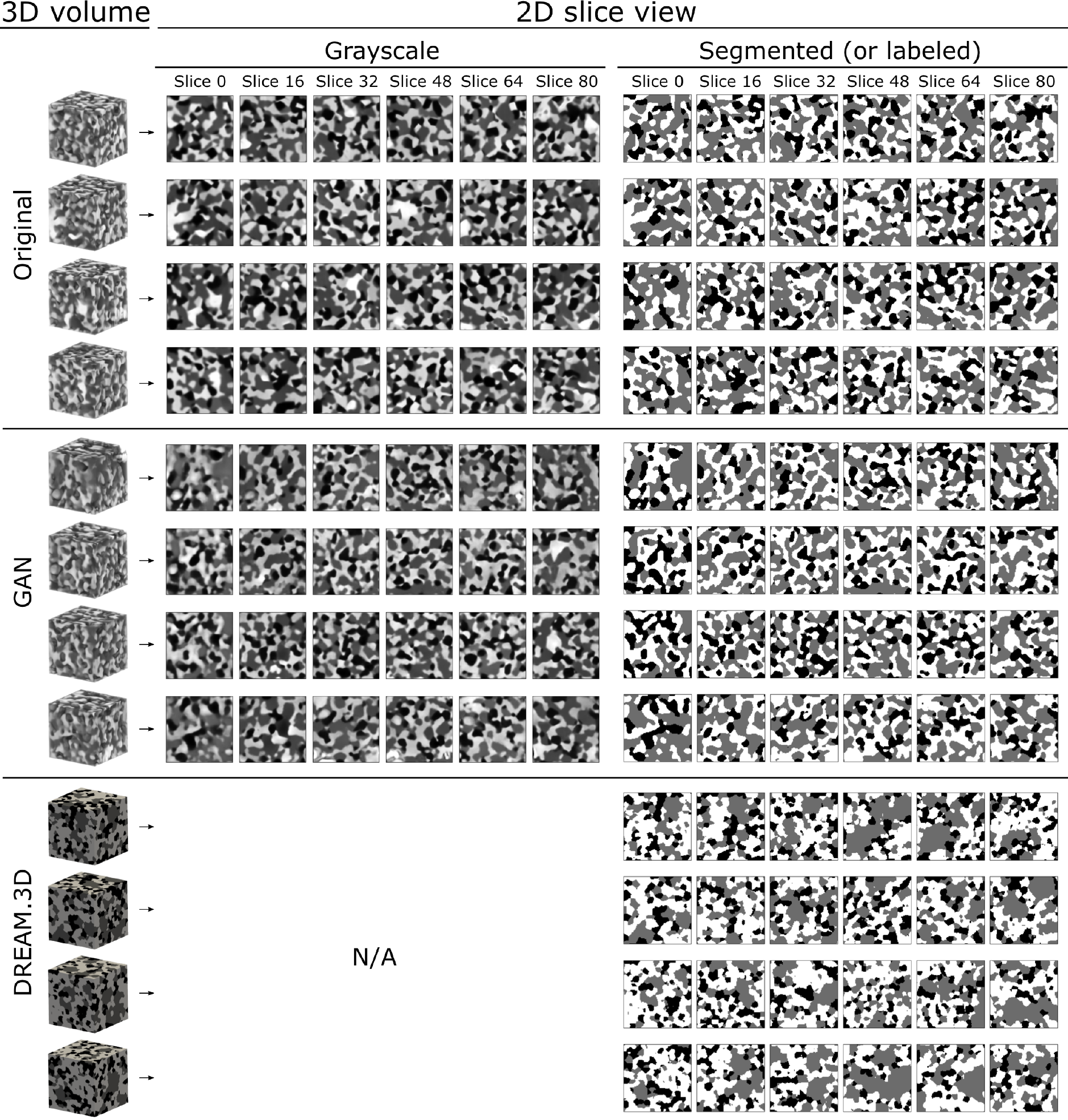}
    \caption{Comparison of original (top), GAN-synthetic (middle), and DREAM.3D-synthetic (bottom) microstructures. Four random volumes (left column) for each category were sampled and 6 2D slices are shown for each, either in grayscale (center column) or segmented/labeled format (right column). The original microstructures consist of three phases: pores (black), nickel (gray) and yttria-stabilized zirconia (bright-gray or white). For the original and GAN structures, segmentation of grayscale images was performed as a post-processing step. DREAM.3D generates microstructures with labeled fields, which are segmented by definition.}
    \label{fig:slice-compare}
\end{figure}

Figure~\ref{fig:slice-compare} offers a deeper visual comparison between the original (top group) and GAN generated (middle group) microstructures. Four volumes of each are shown in the left column, and six slices (2D images) from them are shown as grayscale images in the middle column. The 2D slices clarify that the anode microstructures consist of three grayscale intensities, corresponding to the three component phases: pores (black), nickel (gray), and yttria-stabilized zirconia (bright-gray or white). Similarly, the GAN-generated  microstructures display three grayscale intensities, and the morphology visually resembles that of the original microstructure. Close visual inspection again uncovers some differences. The end slices (e.g., slice 0) of the generated volumes are blurrier. Also, the finer features of the original volumes contain sharper angles, which are less common in the GAN-synthetic volumes. These visual discrepancies indicate that there is still room to improve the GAN implementation, should more precise similarity be required.

In prior work \cite{hsu2018mesoscale, mahbub2020quantitative,hsu2020Moose}, we used an ellipsoid packing method built into the DREAM.3D software to create microstructures approximating the original PFIB datasets. The DREAM.3D microstructures are generated as segmented data, having three voxel values (colors); examples are shown in Figure~\ref{fig:slice-compare} (bottom group). Visually, the GAN-synthetic morphology resembles the original more closely than the DREAM.3D-synthetic morphology does. For example, compared to the DREAM.3D synthetics, the original structures and the GAN synthetics have greater phase connectivity and less agglomeration of the solid (gray) phases. This is not surprising, considering that DREAM.3D input was limited to phase fraction and the first and second moments of the log-normal particle size distribution. While it is possible that the DREAM.3D structures could be improved by iterative optimization or specifying additional microstructural parameters, important aspects of microstructural topology, including clustering and connectivity, are constrained by the ellipsoid packing scheme and are not directly tunable.

\subsection{Statistical Similarity} \label{sec:stats-sim}
\begin{figure}
    \centering
    \addtolength{\leftskip}{-1.5cm}
    \includegraphics[width=1.2\textwidth]{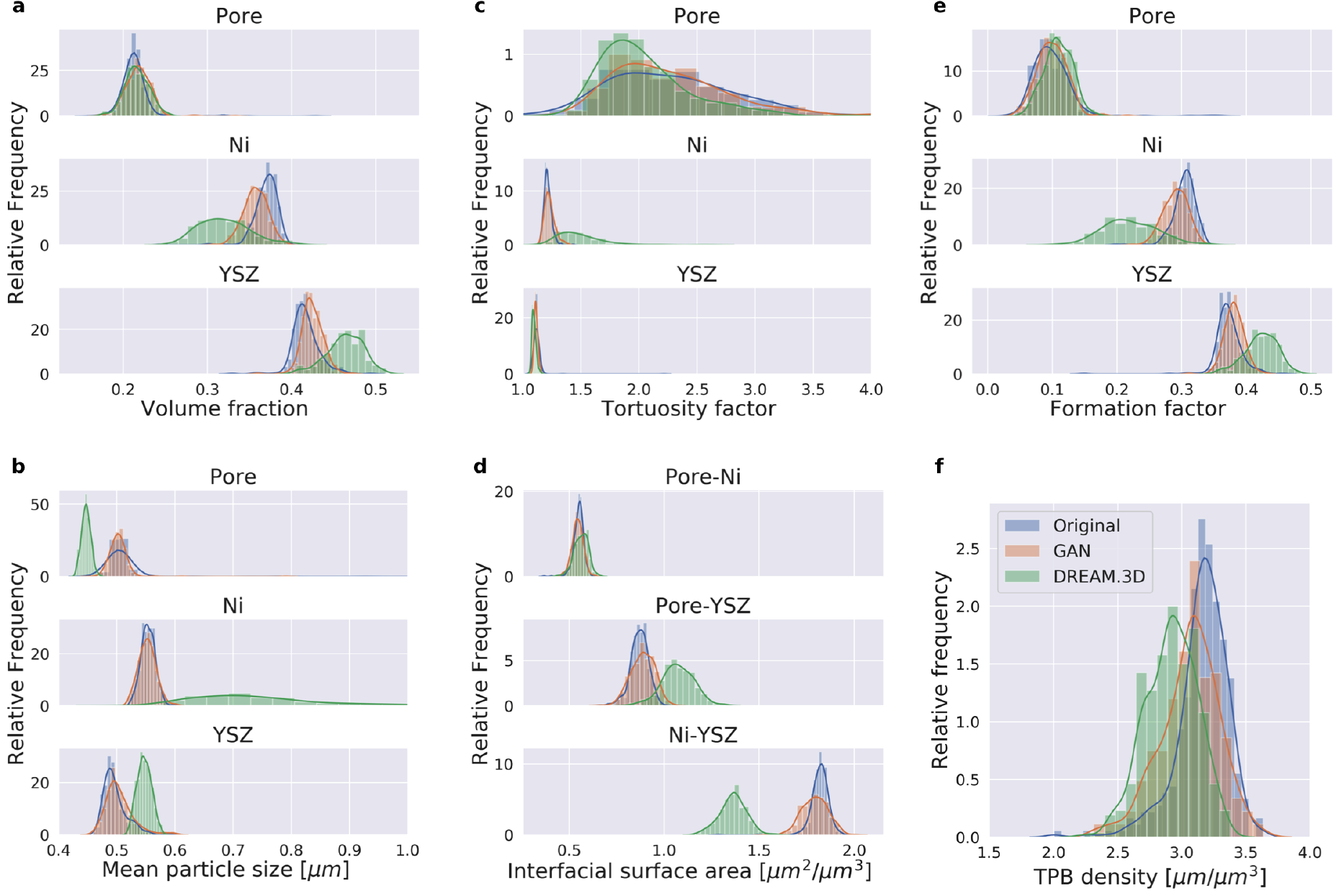}
    \caption{Histograms of original (blue), GAN-synthetic (orange), and DREAM.3D-synthetic (green) microstructure volumes with respect to 3D microstructural metrics: (a) volume fraction of each phase, (b) mean particle size for each phase, (c) tortuosity factor for each phase, (d) interfacial surface area between each pair of phases,  (e) formation factor for each phase, and (f) total triple phase boundary (TPB) density. Phase 1 indicates pores; phase 2 is Ni; and phase 3 is yttria-stabilized zirconia (YSZ). These statistics are based on 323 original (cropped without overlap), 323 GAN-synthetic, and 320 DREAM.3D-synthetic microstructure volumes of the same size. The curves show kernel density estimations.}
    \label{fig:hist-compare}
\end{figure}

The microstructural parameters that are known to impact electrochemical properties in three phase electrodes include volume fractions, particle sizes, tortuosity factors and formation factors (volume fraction / tortuosity factor) for each phase, as well as surface area for each pair of phases, and TPB density. These distributions are shown in Figure~\ref{fig:hist-compare} for the original (blue), GAN-synthetic (orange), and DREAM.3D-synthetic (green) microstructure volumes, and the mean and standard deviation of each distribution is tabulated in Table~\ref{tab:stats-table}. It was shown previously that these parameters can vary significantly across the electrode and between phases \cite{hsu2018mesoscale, mahbub2020quantitative}, as supported by the distributions of the original data in Figure~\ref{fig:hist-compare}. It is impressive that the mean values for the GAN-synthetic volumes match the original volumes within a standard deviation and that there is a very high degree of overlap for each of the 16 distributions shown in Figure~\ref{fig:hist-compare}. The GAN has clearly learned the underlying distributions of features inherent in the original microstructural volumes. 

The DREAM.3D-synthetics, on the other hand, do not match the original data as well as the GAN-synthetics do; for metrics shown in Figure~\ref{fig:hist-compare} (a)-(e), one or more DREAM.3D distribution has significant mismatch or very little overlap with the original distribution. DREAM.3D's methodology targets count-weighted particle statistics rather than volume-weighted statistics presented here, which could contribute to some of the discrepancy seen here. Additionally, the ellipsoid packing construction algorithm constrains the topology of the resulting structures. Although they do not match the original data as well as the GAN-synthetics do, the distributions from the DREAM.3D-synthetics are within a reasonable range relative to those of the original, and have the same relative order of magnitude as those of the original, consistent with past observations \cite{hsu2018mesoscale,mahbub2020quantitative}.

\begin{table}
  \centering
    \caption{Mean $\mu$ and standard deviation $\sigma$ for the distributions of microstructural metrics for the original, GAN generated, and DREAM.3D constructed volumes.}
    \begin{tabular}{c c|c|c|c}
    \hline
    \multicolumn{2}{l |}{Microstructural metric} & \makecell{Original\\$\mu \pm \sigma$} & \makecell{GAN \\ $\mu \pm \sigma$} & \makecell{DREAM.3D \\ $\mu \pm \sigma$} \\
      \hline
\makecell[c]{volume fraction} & \makecell[r]{pore\\Ni\\YSZ} & \makecell{$0.21 \pm 0.02$\\$0.37 \pm 0.01$\\$0.42 \pm 0.02$} & \makecell{$0.22 \pm 0.02$\\$0.36 \pm 0.01$\\$0.42 \pm 0.01$}& \makecell{$0.22 \pm 0.01$\\$0.32 \pm 0.03$\\$0.46 \pm 0.02$}\\
       \hline
\makecell[c]{particle size\\(\si{\um})} & \makecell[r]{pore\\Ni\\YSZ} & \makecell{$0.51 \pm 0.06$\\$0.55 \pm 0.01$\\$0.50 \pm 0.02$} & \makecell{$0.50 \pm 0.02$\\$0.55 \pm 0.01$\\$0.50 \pm 0.02$}& \makecell{$0.45 \pm 0.01$\\$0.74 \pm 0.12$\\$0.55 \pm 0.01$}\\
       \hline
\makecell[c]{tortuosity factor} & \makecell[r]{pore\\Ni\\YSZ} & \makecell{$2.30 \pm 0.61$\\$1.22 \pm 0.03$\\$1.12 \pm 0.06$} & \makecell{$2.29 \pm 0.55$\\$1.23 \pm 0.04$\\$1.11 \pm 0.02$}& \makecell{$2.05 \pm 0.40$\\$1.49 \pm 0.20$\\$1.10 \pm 0.02$}\\
       \hline
\makecell[c]{interfacial area\\(\si{\um^2}/\si{\um^3})} & \makecell[r]{pore-Ni\\pore-YSZ\\Ni-YSZ} & \makecell{$0.55 \pm 0.03$\\$0.87 \pm 0.05$\\$1.82 \pm 0.05$} & \makecell{$0.54 \pm 0.03$\\$0.89 \pm 0.06$\\$1.79 \pm 0.07$}& \makecell{$0.56 \pm 0.04$\\$1.07 \pm 0.09$\\$1.36 \pm 0.07$}\\
       \hline
\makecell[c]{formation factor} & \makecell[r]{pore\\Ni\\YSZ} & \makecell{$0.10 \pm 0.03$\\$0.30 \pm 0.02$\\$0.37 \pm 0.02$} & \makecell{$0.10 \pm 0.02$\\$0.29 \pm 0.02$\\$0.38 \pm 0.02$}& \makecell{$0.11 \pm 0.02$\\$0.22 \pm 0.04$\\$0.42 \pm 0.03$}\\
       \hline
       \multicolumn {2}{c |}{TPB density (\si{\um}/\si{\um^3})} & \makecell{$3.16 \pm 0.20$} & \makecell{$3.06 \pm 0.23$}& \makecell{$2.92 \pm 0.20$}\\
\hline
\end{tabular}
 \label{tab:stats-table}
\end{table}

\begin{figure}
    \centering
    \addtolength{\leftskip}{-1.5cm}
    \includegraphics[width=1.2\textwidth]{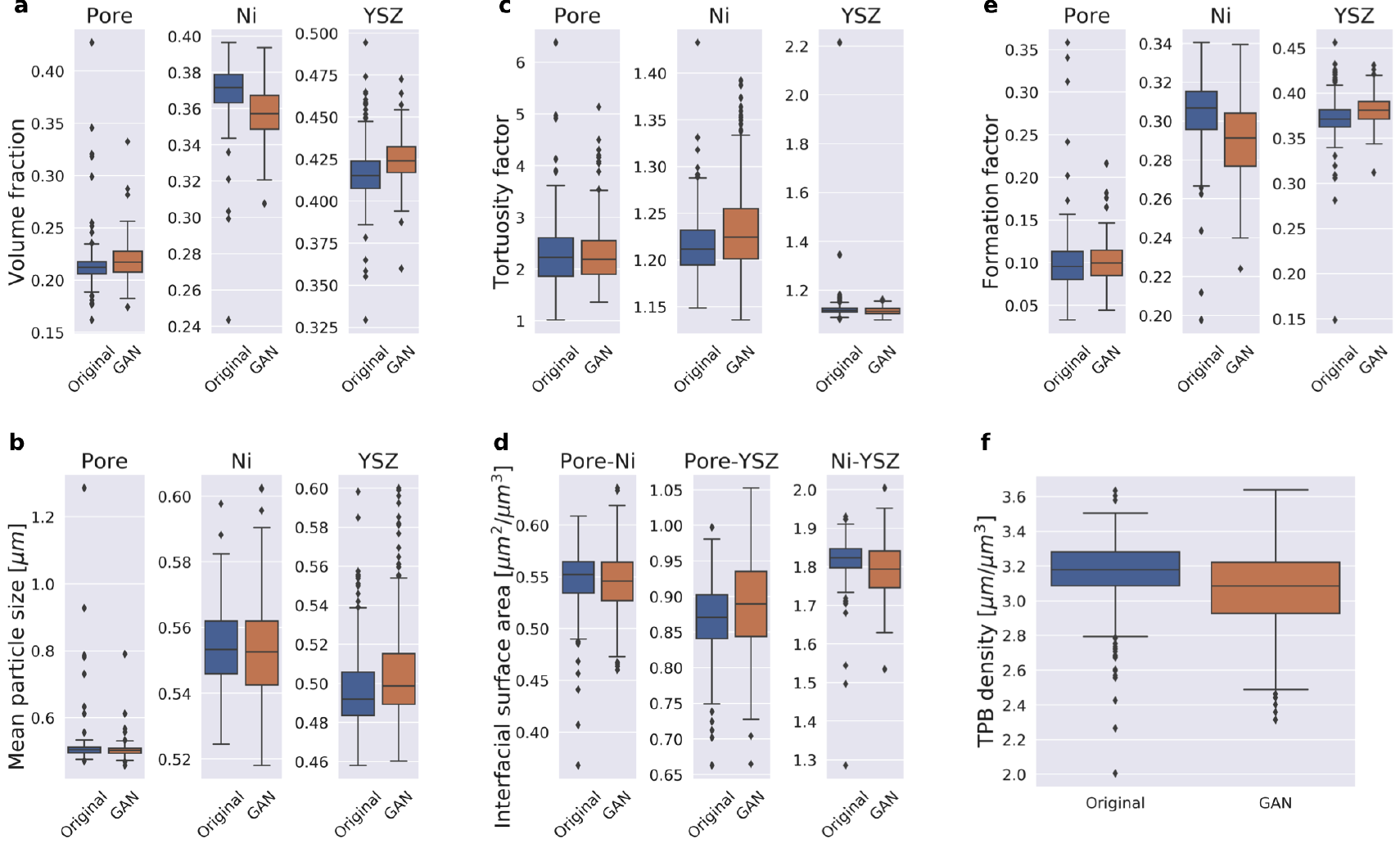}
    \caption{Boxplots of the distributions of microstructural parameters from the original (blue) and GAN-synthetic (orange) microstructure volumes: (a) volume fraction of each phase, (b) mean particle size for each phase, (c) tortuosity factor for each phase, (d) interfacial surface area between each pair of phases,  (e) formation factor for each phase, and (f) total triple phase boundary (TPB) density. Phase 1 indicates pores; phase 2 is Ni; and phase 3 is yttria-stabilized zirconia (YSZ). The inter-quartile range is indicated by the box, the mean by the horizontal line within the box, the minimum ($Q1 - 1.5 \mathrm{IQR}$) and maximum ($Q_3 + 1.5 \mathrm{IQR}$) range by the horizontal lines below and above the box, and outliers are represented by diamonds. $Q_3$ is the upper quartile, $Q_1$ is the lower quartile, and IQR is the interquartile range.}
    \label{fig:box-compare}
\end{figure}

To further evaluate the statistical similarity between the original and GAN microstructures, the distributions in Figure~\ref{fig:hist-compare} are presented as boxplots in Figure~\ref{fig:box-compare}. All pair-wise boxplots have significant overlaps of the interquartile range (IQR), or the box region, as well as the ``minimum''-``maximum'' range (from $Q1 - 1.5 \mathrm{IQR}$ to $Q_3 + 1.5 \mathrm{IQR}$). This reinforces the assertion that the GAN is able to generate microstructures that are statistically representative of the original microstructures, especially around the central portion of the distributions. We define outlier values as those outside of the statistical ``minimum" and ``maximum" values, shown as diamonds in Figure \ref{fig:box-compare}. There are more outliers for the original microstructures than for the GAN-synthetics, and these outliers cover a wider range of values. Thus, while the current GAN implementation captures the central portions of the distributions, further work is necessary to capture the full variability, particularly the outliers, of the original microstructures.

\subsection{Electrochemical Performance}
The microstructure metrics quantified in the previous section have been used in effective medium theory models to describe electrochemical properties or performance of electrodes \cite{Adler1996, Mason2018, Yang2018}. The statistical similarity between the original and GAN structures suggest that their effective medium properties will match as well. For a more direct comparison of local performance, microstructurally resolved, finite element simulations of electrochemistry were carried out on 30 volumes from each microstructure type: original, GAN-synthetic, and DREAM.3D-synthetic. These 90 sub-volumes, which are randomly sampled subsets of the microstructure data in Figure~\ref{fig:hist-compare}, were subjected to a finite element computational workflow described in detail elsewhere \cite{hsu2017towards, hsu2020high, hsu2020Moose}. In essence, the voxelated microstructure images were meshed to produce morphology-conforming finite element computational domains, and a standard reaction-and-transport electrochemistry model was implemented across the meshed microstructures. The simulation model only works on microstructural features fully interconnected to the external source / sink locations of the transport species, and therefore only includes active triple phase boundaries. We note that, though not shown in Figure~\ref{fig:hist-compare}, the original and GAN-synthetics have similar values of inactive TPB densities, but the DREAM.3D has significantly higher inactive TPB densities, highlighting again the difference between GAN-synthetics and DREAM.3D synthetics for capturing locally resolved topological features. 

\begin{figure}
    \centering
    \includegraphics[width=\textwidth]{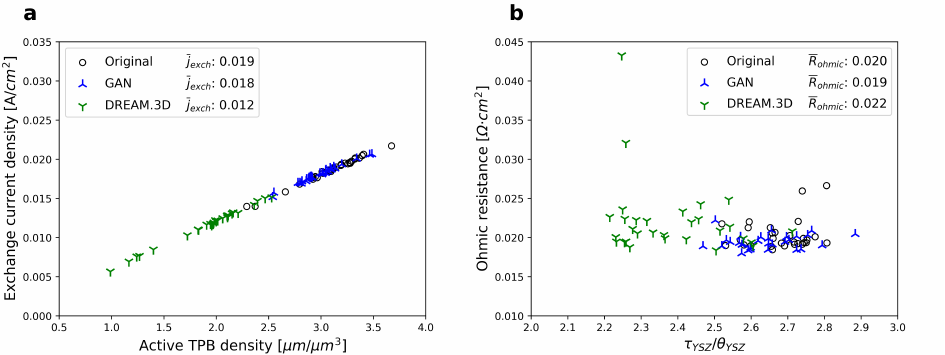}   
    \caption{Simulated electrochemical performance metrics from 30 original, 30 GAN-synthetic, and 30 DREAM.3D-synthetic microstructure volumes. (a) Exchange current density plotted against the triple phase boundary (TPB) density. (b) Ohmic resistance plotted against the inverse formation factor for YSZ, $\tau_{\mathrm{YSZ}} / \theta_{\mathrm{YSZ}}$. For each microstructure category, the mean exchange current density or the mean ohmic resistance is shown in the legend. The microstructure volumes considered here are randomly sampled subsets of the microstructure data for Figure~\ref{fig:hist-compare}. Note that the simulations were operated directly on meshed microstructures, thus retaining topological information.}
    \label{fig:simulated-performance}
\end{figure}

At a given global overpotential $V$, integrating the current $I$ flowing through the electrolyte layer allows one to plot current density versus overpotential for all 90 subvolumes. Those $I-V$ plots can then be fit using a standard activation / ohmic loss model, which yields an effective exchange current density $j_0$ and an effective ohmic resistance $R_{ohmic}$ for each subvolume (see \cite{hsu2020Moose}). As such, the electrochemical performance of each sub-volume can be described using two distinct parameters. For all 90 volumes, the exchange current density is plotted against the active TPB density in Figure~\ref{fig:simulated-performance}(a), and the ohmic resistance is plotted against the inverse of the YSZ formation factor ($\tau_{\mathrm{YSZ}} / \theta_{\mathrm{YSZ}}$) in Figure~\ref{fig:simulated-performance}(b).

Based on commonly used approximations, one might expect the exchange current density to be proportional to TPB density and the effective ohmic resistance to be inversely proportional to formation factor \cite{hsu2020Moose}. Indeed, the exchange current density in Figure~\ref{fig:simulated-performance}(a) is a nearly perfect linear function of active TPB density (it even extrapolates to the origin). As such, the main difference in performance of these electrodes comes from the variability in the active TPB density. Overall, the original and the GAN-synthetic microstructures have highly overlapped scatter distributions. The GAN-synthetics have a slightly lower mean value of active TPB density, and therefore a slightly lower mean value of $j_0$, and the range of data is slightly wider for the original sub-volumes than for the GAN-synthetics. The DREAM.3D-synthetic microstructures, on the other hand, have properties that have little overlap with that of the original microstructures. This arises because the distribution of active TPB density of the DREAM.3D-synthetics differs significantly from the original microstructures.

The ohmic resistance in Figure~\ref{fig:simulated-performance}(b) does not appear to be a strong function of the inverse formation factor of YSZ (the phase through which ions flow and which determines the ohmic resistance). This was shown previously \cite{hsu2020Moose}, and is thought to arise because internal microstructural heterogeneity impacts local ionic transport and, therefore, impacts ohmic resistance values. Nevertheless, the original and the GAN-synthetic microstructures have highly overlapped scatter distributions: the GAN-synthetics being slightly wider in formation factors and the original being wider in the ohmic resistance values. On the other hand, DREAM.3D-synthetic microstructures have little overlap with the original microstructures: the inverse formation factors are lower and wider, while the ohmic resistance values are slightly higher and wider.

Overall, Figure~\ref{fig:simulated-performance} demonstrates that the GAN has learned the underlying distribution of features to an extent that the ensemble electrochemical properties are very well matched with the experimental microstructures. While the DREAM.3D-synthetics are somewhat similar to the originals, they do not do well in matching local connectivity at the sub-volumes sizes investigated here, and therefore result in poorer matches to the performance of the original microstructures. Furthermore, as noted above, the performance simulations reported here presume electrochemical activity is almost entirely restricted to the TPBs. For models that include activity through two-phase boundaries (e.g. through a cathode phase bulk), the performance of the GAN structures would improve further relative to DREAM.3D, since it also captures these features with greater fidelity [Figure~\ref{fig:hist-compare}(d)].

\section{Conclusions}
In this work, we implemented a GAN model for learning and generating microstructural images. This is among the first applications of GAN to 3D microstructure generation at a scale, complexity, and fidelity suitable for ICME applications. We show that the GAN model can generate realistic 3D, topologically complex, multiphase, grayscale microstructures that closely resemble the original, experimental structures in terms of visual appearance, statistical representation of geometric and topological properties, and simulated electrochemical performance. Compared to the commonly-used microstructure generation algorithm DREAM.3D, the GAN results are structurally and electrochemically more realistic. Besides its superior fidelity, we consider that the GAN model has the following merits:
\begin{itemize}
    \item Generality---The GAN framework can be applied to arbitrary materials systems, microstructural morphologies, and imaging modes, making it a general and flexible tool for microstructure generation.
    \item Autonomy---Conventional microstructural construction methods require the user to define a set of metrics that characterizes microstructural geometry and topology. The GAN generates realistic structures independent of user assumptions; that is, it reproduces microstructural metrics without needing to know them.
    \item Throughput---A trained generator can synthesize an arbitrarily large number of unique microstructural volumes.
    \item Computational efficiency---Although training the GAN model requires substantial HPC resources, using it to generate synthetic microstructures is quite efficient.
\end{itemize}

However, we also note that the GAN implementation has the following relative shortcomings:
\begin{itemize}
    \item Training data---Obtaining a sufficient volume of 3D microstructural data to train a GAN via experiments or physical simulation can be costly or challenging. In contrast, conventional construction methods such as DREAM.3D can generate microstructures from statistical descriptors, without requiring 3D image data.
    \item High performance computing---3D image generation requires high model complexity (number of trainable parameters). As such, high performance multi-GPU training is a requirement.  
    \item Synthetic volume---Due to the computational cost of training, this work was limited to generation of 3D images at the scale of $96 \times 96 \times 96$ voxels. Larger volumes require more computational resources in proportion to system size.
    \item Tunability---While conventional construction methods allow users to alter input variables to explore different parameter spaces (e.g. phase fraction, particle size), the present GAN implementation is limited to generating microstructures from the parameter space on which it was trained. However, developing tunable GANs is an active area of research \cite{iyer2019conditional}.
\end{itemize}

Finally, it is worth commenting on the role of this work in the context of applied machine learning models for tasks in the physical sciences. We have shown that based on the metrics discussed in this paper, our GAN model outputs valid and physical microstructures. Thus, a major contribution of our work is the extensive validation of the GAN model output beyond nebulous or subjective measures of visual similarity. This is a necessary step for advancing GAN generative methods as tools supporting ICME and other scientific and engineering endeavors.

\section{Acknowledgements}
This work was performed in support of the U.S. Department of Energy National Energy Technology Laboratory's ongoing research under the RSS contract 89243318CFE000003. EAH acknowledges support by the National Science Foundation under grant CMMI-1826218 and the Air Force D3OM2S Center of Excellence under agreement FA8650-19-2-5209.

\section{Disclaimer}
This work was funded by the Department of Energy, National Energy Technology Laboratory, an agency of the United States Government, through a support contract with Leidos Research Support Team (LRST). Neither the United States Government nor any agency thereof, nor any of their employees, nor LRST, nor any of their employees, makes any warranty, expressed or implied, or assumes any legal liability or responsibility for the accuracy, completeness, or usefulness of any information, apparatus, product, or process disclosed, or represents that its use would not infringe privately owned rights. Reference herein to any specific commercial product, process, or service by trade name, trademark, manufacturer, or otherwise, does not necessarily constitute or imply its endorsement, recommendation, or favoring by the United States Government or any agency thereof. The views and opinions of authors expressed herein do not necessarily state or reflect those of the United States Government or any agency thereof.

\printbibliography

\end{document}